\documentclass[aps,prl,10pt,superscriptaddress,twocolumn]{revtex4-1}

\usepackage[cmex10]{amsmath}
\usepackage{amssymb}

\usepackage{graphicx}
\usepackage{braket}
\usepackage{natbib}

\usepackage{times}
\usepackage{siunitx}
\usepackage{color}
\usepackage{comment}

\begin{document}
\title{\textcolor{black}{Pseudo-2D superconducting quantum computing circuit for the surface code:\\ the proposal and preliminary tests}}
\author{H. Mukai}
\affiliation{Department of Physics, Tokyo University of Science, 1--3 Kagurazaka, Shinjuku, Tokyo 162--0825, Japan}
\affiliation{RIKEN Center for Emergent Matter Science (CEMS), 2--1 Hirosawa, Wako, Saitama 351--0198, Japan}
\author{K. Sakata}
\affiliation{Department of Physics, Tokyo University of Science, 1--3 Kagurazaka, Shinjuku, Tokyo 162--0825, Japan}
\affiliation{RIKEN Center for Emergent Matter Science (CEMS), 2--1 Hirosawa, Wako, Saitama 351--0198, Japan}
\author{S. J. Devitt}
\affiliation{RIKEN Center for Emergent Matter Science (CEMS), 2--1 Hirosawa, Wako, Saitama 351--0198, Japan}
\affiliation{Centre for Quantum Software \& Information (QSI), Faculty of Engineering \& Information Technology, University of Technology Sydney, Sydney, NSW, 2007, Australia}
\author{R. Wang}
\affiliation{Department of Physics, Tokyo University of Science, 1--3 Kagurazaka, Shinjuku, Tokyo 
162--0825, Japan}
\affiliation{Centre for Quantum Software \& Information (QSI), Faculty of Engineering \& Information Technology, University of Technology Sydney, Sydney, NSW, 2007, Australia}
\author{Y. Zhou}
\affiliation{RIKEN Center for Emergent Matter Science (CEMS), 2--1 Hirosawa, Wako, Saitama 351--0198, Japan}
\author{Y. Nakajima}
\affiliation{Department of Physics, Tokyo University of Science, 1--3 Kagurazaka, Shinjuku, Tokyo 162--0825, Japan}
\affiliation{Centre for Quantum Software \& Information (QSI), Faculty of Engineering \& Information Technology, University of Technology Sydney, Sydney, NSW, 2007, Australia}
\author{J. S. Tsai}
\email{e-mail: tsai@riken.jp}
\affiliation{Department of Physics, Tokyo University of Science, 1--3 Kagurazaka, Shinjuku, Tokyo 162--0825, Japan}
\affiliation{RIKEN Center for Emergent Matter Science (CEMS), 2--1 Hirosawa, Wako, Saitama 351--0198, Japan}


\maketitle

\textbf{Of the many potential hardware platforms, superconducting quantum circuits have become the leading contender for constructing a scalable quantum computing system. 
All current architecture designs necessitate a 2D arrangement of superconducting qubits with nearest neighbour interactions, compatible with powerful quantum error correction using the surface code. 
A major hurdle for scalability in superconducting systems is the so called wiring problem, where qubits internal to a chip-set become inaccessible for external control/readout lines. 
Current approaches resort to intricate and exotic 3D wiring and packaging technology which is a significant engineering challenge to realize, while maintaining qubit fidelity. 
Here we solve this problem and present a modified superconducting micro-architecture that does not require any 3D external line technology and reverts back to a completely planar design. 
This is enabled by a new pseudo-2D resonator network that provides inter-qubit connections via airbridges. 
We carried out experiments to examine the feasibility of the newly introduced airbridge component. 
\textcolor{black}{Our simulation shows the measured quality factor of the airbridge-resonator are below the threshold for surface code and it does not limit gate fidelity. 
The measured crosstalk between crossed resonators is $\SI{-49}{dB}$ at most on resonant}. 
\textcolor{black}{The spatial separation between the external wirings and the inter-qubit connections would result with a relatively limited crosstalk between them that would not increase as the size of the chip-set increases}. 
\textcolor{black}{This architecture indicates the possibility that a large-scale, fully error corrected quantum computer could be constructed by monolithic integration technologies without additional overhead and without special packaging know-hows}. 
}

Recently, architecture designs for large-scale quantum computers are becoming more and more comprehensive. 
This field frequently includes a large amount of quantum engineering specifying how qubits will be manufactured, controlled, characterized and packaged in a modular manner for fault-tolerant, error corrected quantum computation~\cite{Jones:2012aa,Nemoto:2014aa,Gimeno-Segovia:2015aa,Hill:2015aa,Lekitsch:2017aa}. 
The vast majority of architectures base their designs on the surface code because it has one of the highest fault-tolerant thresholds of any error correction code, easing the physical fidelity requirements on the hardware. 

Superconducting quantum circuits have emerged as a major contender for a scalable hardware model for the surface code. 
Superconducting qubits are fabricated with inter-qubit wirings for nearest neighbor interactions and each individual qubit requires external physical access such as bias lines, control lines, and measurement devices. 
However, as the 2D array is scaled up, planar accessibility for control lines become a problem. 
Such challenges are sometime referred to as the wiring problem, where physical qubits in the interior are no longer accessible, in plane, from the edge~\cite{Gambetta:2017aa}. 

Compared with classical silicon integrated circuits, it is much more difficult to achieve such wiring in superconducting quantum circuits. 
To individually access every qubit in the 2D qubit array, standard multi-layer wiring technologies for silicon integrated circuits simply cannot be embraced as it generally requires the introduction of decoherence enhancing, low quality inter-layer insulators. 
Therefore, in current superconducting systems, many groups are forced to utilize non-monolithic bulky 3D wiring technologies (see Fig.~1), such as flip-chip bonding, pogo pin and through silicon via (TSV)~\cite{Barends:2014aa,Takita:2017aa,Reagoreaao3603,Chou:2018aa, Bejanin:2016,Vahidpour:2017,Foxen:2018aa,rosenberg_3d_2017}. 

\begin{figure*}[!t]
  \includegraphics[width=18cm]{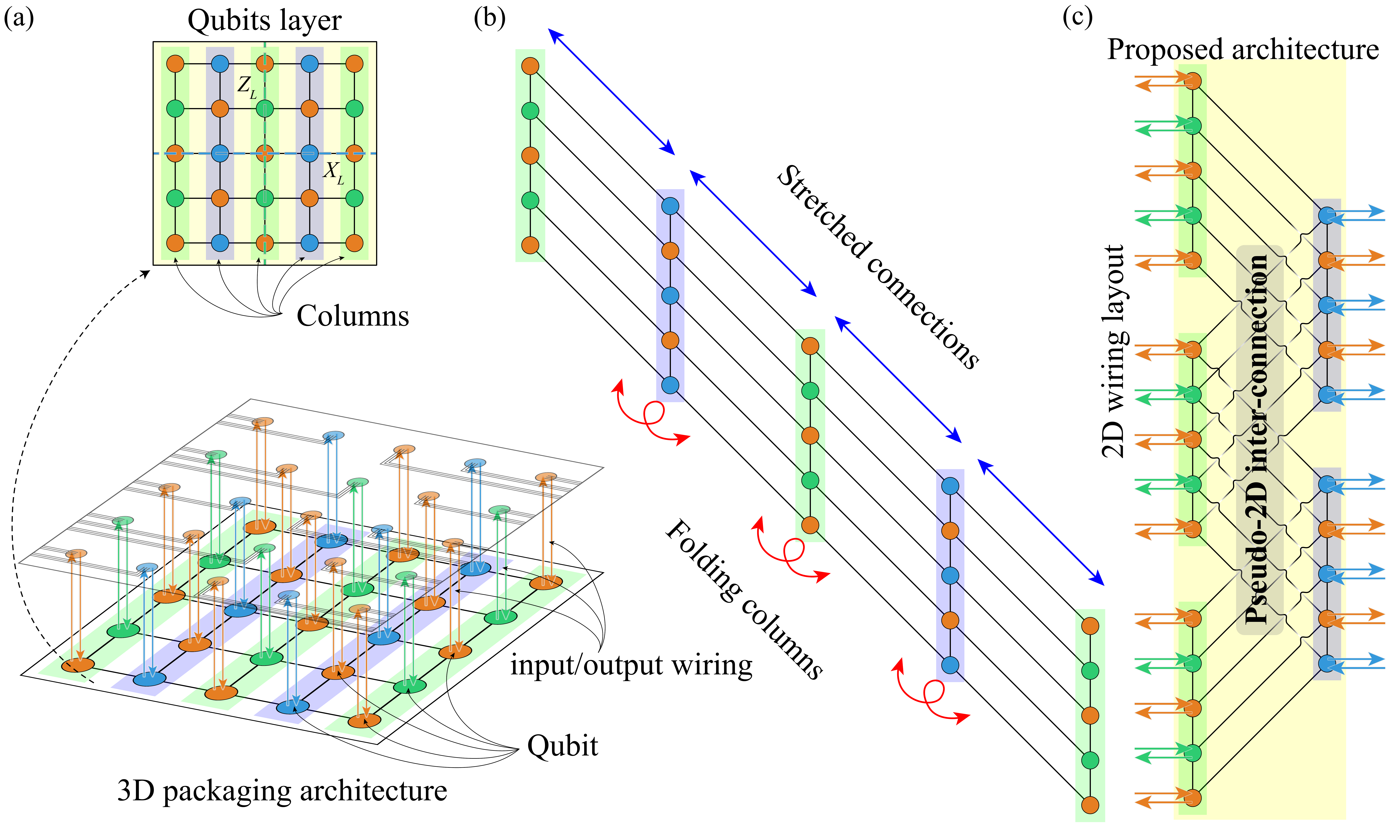}
  \caption{
  Standard circuit architecture and our proposed architecture for surface codes with $5\times 5$ qubit array. 
  (a) The standard system utilizing global multi-layer structures, a separated qubit layer (shown top item) and a control/readout layer (clear layer in bottom item). 
  Qubits are divided into data qubits (orange circle indicated) and X (blue circle indicated), Z (green circle indicated) syndrome qubits, and all nearest neighbor qubits are connected by inter-qubit wiring. 
  Vertical arrows indicate input/output wirings. 
  (b) The folding operation of proposed architecture. 
  In a horizontal direction, inter-connection of 2D qubit array are stretched out, \textcolor{black}{while maintaining the resonator frequency. }
  At each qubit column, the inter-connection is folded. 
  (c) The proposed planar architecture for surface code. 
  After process (b), out architecture have completely planar external wirings (all arrows do not intervene any wiring, external lines and also inter-qubit connections) with a help of pseudo-2D inter-connections. 
  }
  \label{fig:architecture}
\end{figure*}

Our new architecture for the surface code is obtained by transforming the 2D qubit array to a bi-linear array. 
Fig.~\ref{fig:architecture} shows the mapping between before and after the transformation. 
The square lattice Fig.~\ref{fig:architecture}(a) is divided into many columns. 
Next, the connections between columns are stretched [Fig.~\ref{fig:architecture}(b)], and then, the columns are folded on top of each other successively, as shown in Fig.~\ref{fig:architecture}(c). 
\textcolor{black}{As the connections are stretched out, frequencies of resonators are maintained. 
Therefore, both circuits before and after transformation occupy approximately the same area, as shown in the yellow colored areas in Fig.~\ref{fig:architecture}(a) and (c).
}
The resulted equivalent surface code circuit is a bi-linear array of the original 2D structure. 

The folding operations liberate the columns locked deeply inside the original 2D lattice and brings them out to the edges of the bi-linear array. 
Therefore the external control/readout lines connected to each qubit are accessible from the edges of the chip. 
This novel arrangement allows all these external connection to be prepared in a completely standard 2D layout. 

The advantage gained in the external wiring by the transformation, however, takes a small toll in the inter-qubit wiring between columns. 
These inter-qubit connections between neighboring columns require multi-level crossings. 
Nonetheless, these 3D structures only need to locally hop over inter-qubit connection lines. 
Thus, the cross-connections between the columns can be described as pseudo-2D. 

In comparison, for the original surface code architecture, the multi-layer wiring grid involves an inter-qubit connection layer and an input/output wiring layer. 
Therefore, a global multi-layer structure, as shown in Fig.~\ref{fig:architecture}(a), is often adopted, utilizing non-monolithic bulky 3D wiring technologies as mentioned earlier. 
Compared to the standard surface code arrangement, the new architecture  has the following obvious advantages:
\begin{enumerate}
\renewcommand{\labelenumi}{(\arabic{enumi})}
\item \textcolor{black}{The complete separation of the input/output wirings and inter-qubit wirings would probably help to suppress crosstalk between external lines and qubits as well as that between external lines and inter-qubit connection lines. Therefore, it is possible that undesired decoherence of qubits due to external wiring would also be reduced.}

\item 2D planar layout of the input/output wirings. 
These wirings, connecting qubits to external electronics and can be constructed by utilizing the standard 2D wide-band (microwave) wiring technology. 
Superconducting resonators for the readout of the qubit can also be prepared with the standard 2D coplanar design.  
\item Local 3D (psuedo 2D) wiring. 
The ends of the inter-qubit connection lines always end up on the same qubit layer, no matter how many 3D hops are involved in the connection. 
In such case, the multi-layer crossing for the new architecture could be realized simply by local monolithic 3D structures, such as superconducting airbridges. 
\end{enumerate}

Moreover, the original square lattice architecture could adopt the local 3D structure (airbriges) for the wire crossings between input/output and inter-qubit connections. 
However, compared with the new architecture, such arrangement would produce strong crosstalk between external wirings and inter-qubit connection lines (cf. point (1) above). 

Consequently, this architecture straightforwardly solves the demanding 3D external wiring problem. 
As already mentioned, a convenient technology to realize the cross wiring is an airbridge; a monolithic microstructure, developed as a low-loss wiring for superconducting qubits that can be fabricated in several ways, including a well-established standard fabrication process~\cite{chen_fabrication_2014,dunsworth_method_2018}. 
A large number of airbridges, compared with the number additionally required for this proposal, are always needed to maintain the uniform ground potential for all coplanar waveguide-based architectures. 

\begin{figure*}[!ht]
  \includegraphics[width=17.6cm]{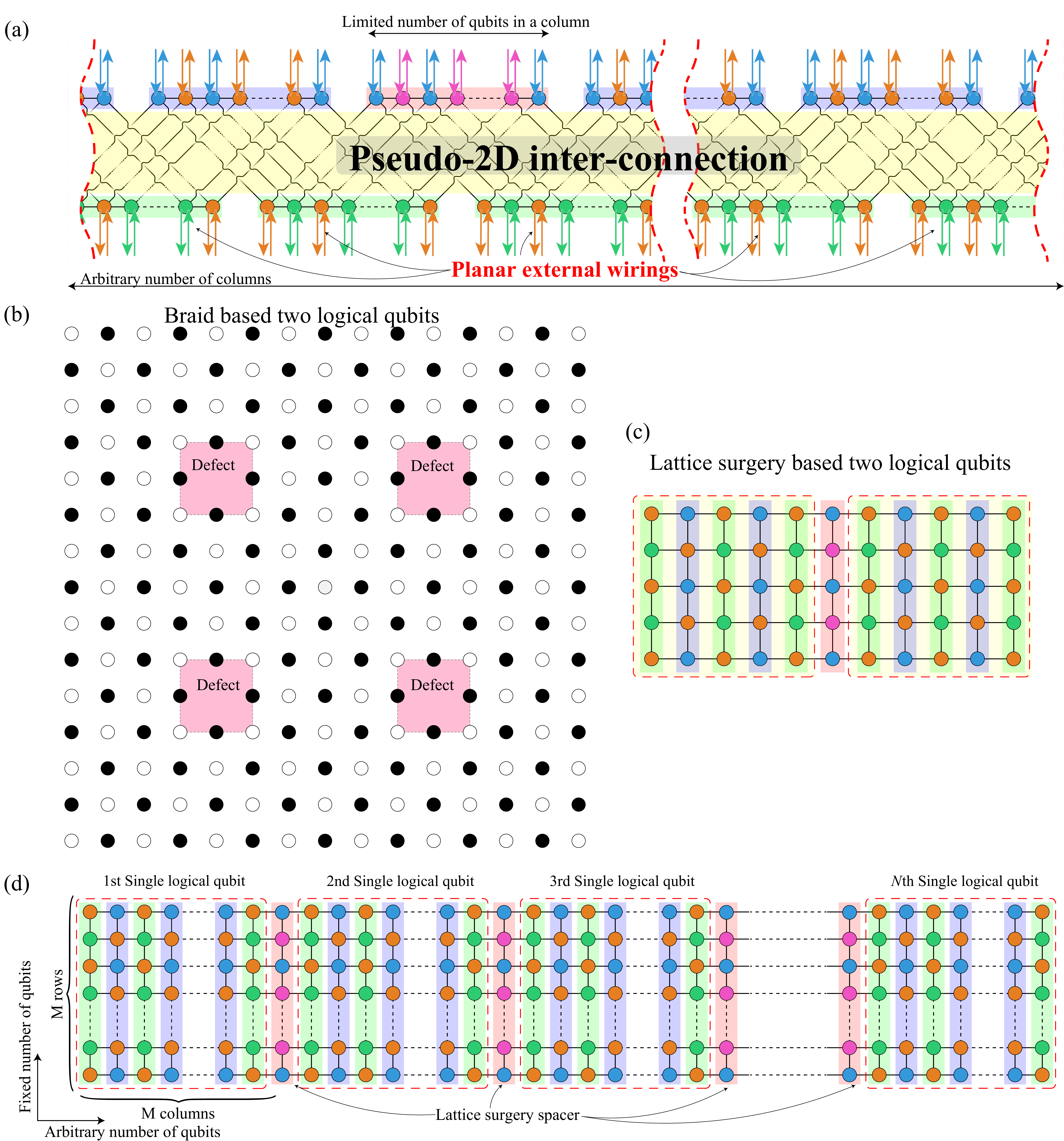}
  \caption{
   (a) Here we illustrate the physical layout of the new architecture. 
   An arbitrarily long, but fixed width surface code can be created using a bi-linear arrangement of superconducting qubits that are cross coupled with airbridged resonators. 
   The fixed width of the surface code ensures that airbridged resonators have a finite length and number of bridged crossings. 
   Each superconducting qubit can be accessed in the plane for control, initialisation and readout technology. 
   \textcolor{black}{
   (b) The figure is a standard braid based arrangement of the surface code, sufficient for encoding two logical qubits of information with a distance $d=3$ surface code.  
   (c) The figure is a standard arrangement of lattice surgery based two square patch, sufficient for encoding two logical qubits of information with a distance $d=3$ surface code.
   }
   (d) The new logical qubit layout consists of square patches of surface code, each encoding a single logical qubit of information.  
   Between square patches there are spacer regions (red column) to allow for lattice surgery based logic operations. 
   This layout maintains a small, fixed width of the physical lattice and converts the computer into a LNN logical qubit array. 
   A technique for logical compilation and operation could include a single extra row of physical qubits stretching the length of the computer to enact a new data bus technique for logic operations using planar codes and lattice surgery~\cite{Paler:2019}.
   }
  \label{fig:logical}
\end{figure*}

To scale up the degree of integration, one needs to consider that increasing the number of qubits $M$ in a column, which is represented as green, blue and red columns in a scaled-up structure of our proposed architecture [Fig.~\ref{fig:logical}(a)], would result with \textcolor{black}{the growth of the required number of airbridges.}
Therefore, one should limit $M$ to a minimally required number for a surface code based computer in effective 2D array.  
This is the arrangement before the transformation shown in Fig.~\ref{fig:logical}(d)]. 
This limitation posed by the number of airbridges results in a subtle change in the design, compared with the standard 2D array for a surface code architecture. 

Typical logical structure of the computer shown in Fig.~\ref{fig:logical}(b) is a 2D array of qubits used for surface code computing, utilizing braid based logic~\cite{Fowler:2012}. 
Logical information is introduced by strategically switching on/off parts of the array to create and manipulate defects, which encode the logical qubits within the computer. 
The larger the 2D array at the physical layer, the more defects can be introduced for number of logically encoded qubits in the computer or the larger each defect can be for the strength of the error correction.  
Logic operations are then performed by topological braiding of the defects around each other. 
In Fig.~\ref{fig:logical}(b), we illustrate a lattice that encodes two logical qubits via four pairs of defects introduced into the lattice (shaded regions), where two pairs are for each logical qubit. 
The defects are encoded using a $d=3$ surface code, which can correct for an arbitrary single qubit error on either of the two encoded defect based qubits. 
In order to realize this \textcolor{black}{defect-based} structure, without significantly compromising the ability to efficiently enact arbitrary error-corrected circuits, \textcolor{black}{the arbitrarily scaling up is required in 2-dimensions.}

In our new design, however, the length of columns in the effective 2D array is limited - due to the number of airbridged crossings in a inter-qubit connection - but an arbitrary number of columns is allowed. 
Therefore, we envisage that lattice surgery encoded logic will be used instead of braid based logic \textcolor{black}{(shown in Fig.~\ref{fig:logical}(c) for the $d=3$ surface code)~\cite{Horsman:2012aa}. 
The lattice surgery encoded logic also can aid the realization of sufficiently fast classical error correction decoding~\cite{Dev09,Fow12+}.}
In lattice surgery, isolated square patches of the planar code ({\em single logical qubit}, which is a surface code analogue that can encode a single piece of logical information) are interacted along a boundary to enact multi-qubit logic gates. 
This reduces the overall physical resource cost of each logical qubit and several results now suggest that lattice surgery techniques will always be more resource efficient when implementing large-scale algorithms~\cite{Herr:2017aa,Litinski2018latticesurgery,Fowler:2018}. 

For a single logical qubit encoded with the planar code, a square 2D array of physical qubits is needed. 
For a quantum code with a distance $d$, a $(2d-1)\times (2d-1)$ array of physical qubits is sufficient, the number of which can be reduced further utilizing rotated planar lattices~\textcolor{black}{\cite{Horsman:2012aa, Litinski2018latticesurgery}} \textcolor{black}{(see methods section)}. 
This results in a Linear Nearest Neighbour (LNN) logical layout of encoded qubits \textcolor{black}{[shown in Fig~\ref{fig:logical}(d)]}, requiring less physical resources than defect-based logical qubits. 
In Fig.~\ref{fig:logical}(d), you can see that there is additional columns of physical qubits only (red colors) that are spacers between each encoded qubit that is required to perform the lattice surgery operations. 

It should be noted that the current methods for circuit compilation using lattice surgery still assumes a 2D nearest neighbor arrangement of logically encoded qubits~\cite{Herr:2017aa,Fowler:2018,Litinski2018latticesurgery}. 
This is because lattice surgery has two basis classes of operations (merges and splits) over two types of boundaries for each planar code qubit (what are known as rough and smooth). 
As merge and split operations can only occur on the single boundary between logical qubit regions, we need to be able to convert between smooth and rough boundaries (which was detailed in Ref.~\cite{Horsman:2012aa}) and hence compilation into this LNN logical structure using pseudo-2D physical qubits layout will require some slight modifications over current techniques~\cite{Herr:2017aa,Litinski2018latticesurgery}. 
However, recent results \textcolor{black}{which} introduce a single additional row of physical qubits to act as a data bus for logic operations can be used and is completely compatible with a LNN arrangement of qubits at the logical level~\cite{Paler:2019}.  

For a large error-correcting code, distance $d$ can be of the order of $d=15-21$ (capable of correcting up to 7-10 errors per logical qubit). 
With a distance $d$ code requiring an array containing $M=2d-1$ rows of qubits with $M=2d-1$ columns, per single logically encoded qubit. 
Consequently, for a quantum computer containing $N$ logical qubits on the planar code, we would utilize an array of $M\times [NM+(N-1)]$. 
Here, first $M$ is the number of qubits in a column, and $NM$ is the number of columns in the array for $N$ logical qubits and the extra factor of $(N-1)$ is the spacing region between each logical qubit needed for the lattice surgery (or a bus system~\cite{Paler:2019}). 
This would translate into a bi-linear array, as shown in Fig.~\ref{fig:logical}(a) of $2\times \frac{1}{2}(2d-1)(2dN-1)$. 
A number of crossing points by inter-connections are at most half the number of qubits in a column, at most \textcolor{black}{$\lceil [(2d-1)-1]/2\rceil = d-1$}, representing the number of airbridges per resonator. 
The factor of $1/2$ comes about due to the fact that alternate resonators (inter-connections) are shared by two qubits. 
Hence, while the number of columns $NM$ linearly increases with the number of logical qubits, the number of airbridges contained in resonator will only be half number of qubits in a column (which is fixed for a given code distance, $d$). 

In practice, the width of this array is related to the number of logical qubits while its length is given by the distance of the planar code used to encode each logical qubit. 
For a heavily error corrected logical qubit, $d = 15\text{--}21$, the total number of qubits in a column will be $M = 29\text{--}41$ with a maximum number of airbridges for a given resonator of $\textcolor{black}{14}\text{--}20$. 
By utilizing planar code encoding and lattice surgery~\cite{Horsman:2012aa} for fault-tolerant logic, we can define our computer as a long, rectangular structure consisting of a LNN array of {\em logical} qubits (requiring compilation of the high level quantum algorithm with LNN constraints~\cite{Paler:2017aa,Herr:2017aa,Paler:2019}).

\textcolor{black}{
To make a feasibility study of this new circuit scheme, we carried out preliminary evaluations of its most important new- \\ \\ \\ \\ ly introduced component, namely, the pseudo-2D inter-connection consists of crossed resonators with airbridges. 
We studied the dependence of the gate fidelity on the quality factor of resonators, of which the center line contains airbridges. 
We also studied the crosstalk between crossed coplanar resonators in the pseudo-2D inter-connection network.} 

Examining if airbridges can be used while still satisfying the error requirements for surface code quantum error correction, we carried out both experimental and numerical tests on a system containing a lossy resonator for connecting two qubits. 
Usual research in superconducting quantum circuit employ a very lossless resonator to eliminate its contribution. 
However there is little research related the dependence on the resonator quality factor. 
Therefore, the numerical test reveals a lower limit of the internal quality factor, and the experimental test illustrates the possibility that this proposed architecture is expected to be viable using current technology, without special 3D techniques. 

\begin{figure}[!t]
  \includegraphics{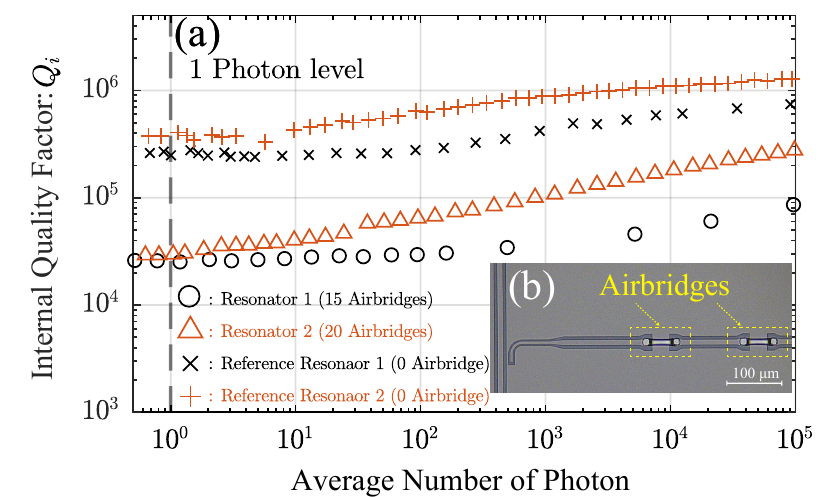}
  \caption{
  The measured internal quality factors versus average photon number in resonators (a). 
  $\bigcirc$ and $\triangle$ indicate datasets of coplanar resonators with 15 and 20 airbridges at center conducting line. 
  $\times$ and $+$ indicate datasets of reference coplanar resonators with no airbridge for $\bigcirc$ and $\triangle$ resonators, respectively. 
  We fabricated two chips, on which $\bigcirc$ and $\times$ resonators are fabricated on one chip, $\triangle$ and $+$ are on the other chip. 
  The dashed line correspond to one average photon level. 
  The $Q_i$ are fitted by standard methods~\cite{probst_efficient_2015}. 
  The detailed structure of the airbridges showing a continuous ground plane below, forming micro-strip like structures locally (b).
  \textcolor{black}{
  Coupling Quality factor $Q_c$ and frequency of each resonator $\omega_r/2\pi$;
  $(Q_c,\,\omega_r/2\pi) = 
  \{
  \bigcirc: (3.141\times 10^5,\,\SI{10.1326}{GHz}),\,
  \triangle:(5.273\times 10^5,\,\SI{7.80465}{GHz}),\,
  \times:   (3.959\times 10^5,\,\SI{9.43147}{GHz}),\,
  +:        (5.162\times 10^5,\,\SI{7.23419}{GHz})\}$
  }
  }
  \label{fig:qfactor}
\end{figure}

We prepared chips using a standard fabrication method for the airbridges, each of which contain both a resonator with airbridges and a reference resonator \textcolor{black}{made out of $\SI{50}{nm}$ thick Nb film}. 
Airbridge design of chips including the interval of airbridge position is identical. 
Any difference in fabrication is only related to the number of airbridge (15 and 20). 
Each wafer were treated under the same conditions, but they were not fabricated at the same time. 
Fig.~\ref{fig:qfactor}(a) shows the measured internal quality factor $Q_i$ of resonators containing \textcolor{black}{15 (black symbol) and 20 (red symbol)} airbridges in the center conducting lines [Fig.~\ref{fig:qfactor}(b)], with reference resonators also illustrated \textcolor{black}{in Fig.~\ref{fig:qfchip} (see method section)}. 
The quality factor of the resonator with airbridges at the center line, are $> 2.3\times 10^4$ at the power of the single photon level. 
In comparison with the reference coplanar resonators, which does not have airbridges, the quality factor decreased by about one order of magnitude. 
The quality factor of resonator with 20 airbridges is higher than one with 15 airbridges. 
The result of this reversal between 15 and 20 may be due to deviations that occur in the fabrication process. 

\begin{figure}[!t]
  \includegraphics{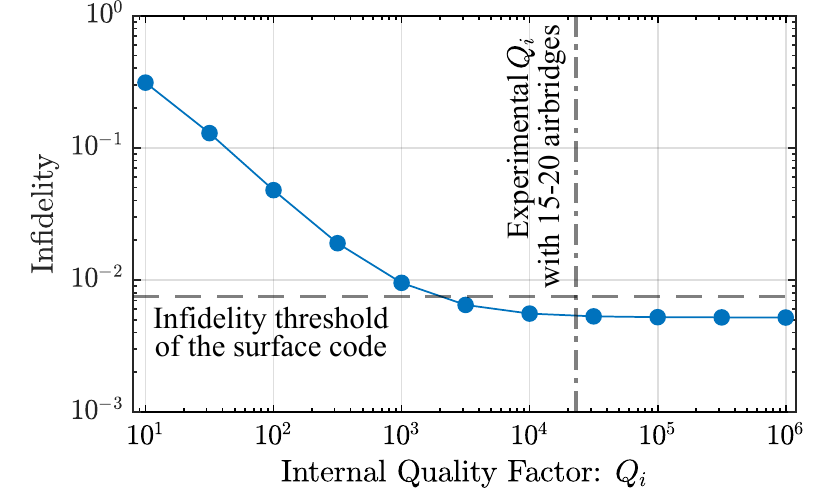}
  \caption{The simulated average gate infidelity of the CZ gate, via the resonator, versus the quality factor of the resonator. 
  \textcolor{black}{Frequency of the {\it i}th qubit between ground and first excited levels: $\omega_i^{01}/2\pi=\SIlist{5.6;5.8}{GHz}$};  
  anharmonicity of the {\it i}th qubit: $\eta_i/2\pi=\SI{-200}{MHz}$; 
  resonator frequency:  \textcolor{black}{$\omega_r/2\pi=\SI{6}{GHz}$}; 
  the coupling constant between the {\it i}th qubit and resonator: $g_i/2\pi=\SI{81.2}{MHz}$;
  \textcolor{black}{
  the effective coupling strength between qubits: $g_{eff}/2\pi=\SI{3}{MHz}$; and
  the gate time; $\SI{117.9}{ns}$}.
  Dash line shows the threshold of surface code. 
  Dash-dotted line indicates the experimental $Q_i$ with $15\text{--}20$ airbridges.}
  \label{fig:fidelity}
\end{figure}

To appraise the effect of extra loss resulting from the insertion of airbridges, we simulated an average gate infidelity of a CZ gate in our system, where two \textcolor{black}{transmon-type qubits} are coupled through a damped (lossy) resonator. 
Here each qubit has three energy levels, anharmonicity $\eta_i$, the resonator has five energy levels with photon leakage rate ($\kappa_i = \omega_r/Q_i$), \textcolor{black}{and the coupling constant between each qubit and resonator $g_i$. In the system, we ignored the qubit-qubit direct coupling. In the system, we ignored the qubit-qubit direct coupling.} 
We adjusted the state of the system to the condition for CZ gate which is that the energy difference from ground to first \textcolor{black}{excited} levels on one qubit is the same as the energy difference from first to second \textcolor{black}{excited} levels on the other qubit.
Then, we calculated the time evolution of this system, and finally got the average gate fidelity $F$. 
\textcolor{black}{
In this simulation, at the beginning, the system was set to the CZ gate condition, this means that simulation started after rising of gate pulse. 
In other words, we simplified simulation by ignoring effects of practical pulse shape. 
The leakage from the total system to external environment are also assumed that the resonator is responsible. 
These assumptions are made to evaluate the dependency of fidelity on quality factor of resonators.
}
Fig.~\ref{fig:fidelity} is the result of the simulation, showing the infidelity dependence on the quality factor of the resonator $Q_i$. 
The result indicates that the required $Q_i$ for the infidelity threshold of the surface code ($1-F<\SI{0.75}{\percent}$) is $2 \times 10^{3}$, and the infidelity is saturated at $Q_i>10^4$. 

The experimental internal quality factor of a resonator with airbridges at centerline one order of magnitude greater than what is required by our simulations. 
In this experiment, current existing technologies were used. 
Therefore, this results strongly suggests that our proposed system, with real parameters, is feasible. 

\begin{figure}[!t]
  \includegraphics[width=8.4cm]{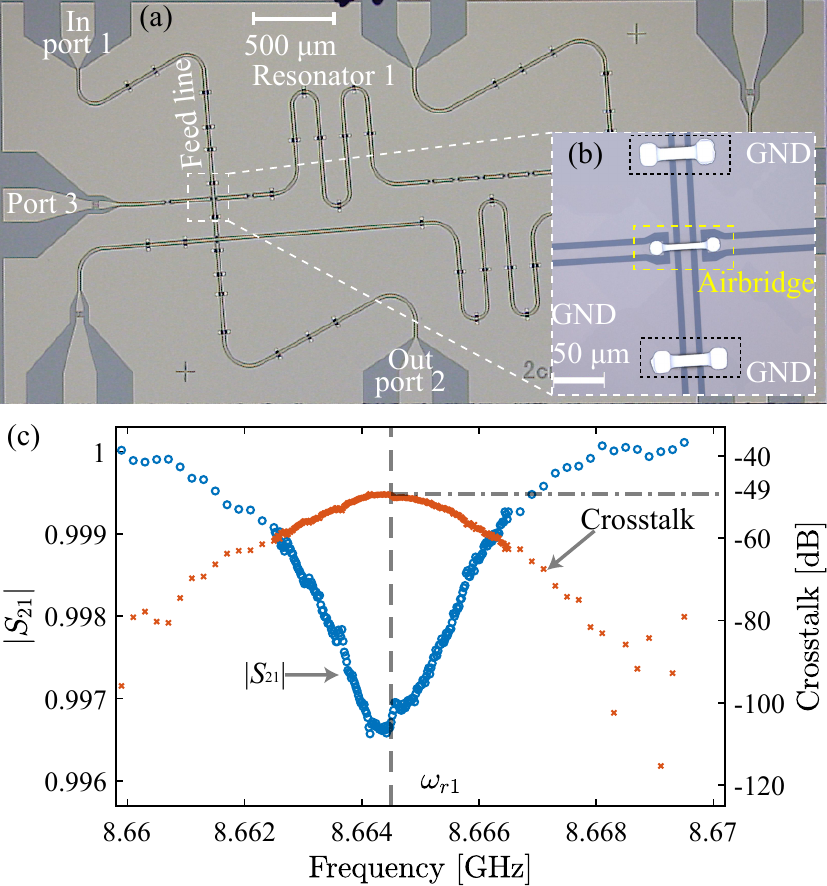}
  \caption{
  (a) Optical image of the chip for crosstalk measurements. 
  \textcolor{black}{Two horizontal lines in parallel are half-wavelength resonators}. 
  Two vertical lines in parallel are feed lines to measure coupling strengths to the resonators at cross point via airbridges. 
  (b) Detail image of the cross point utilizing an airbridge. 
  \textcolor{black}{The center airbridge connects left to right signal lines of resonator over the vertical feed line. 
  Top and bottom airbridges connects ground (GND) planes, which are separated}.
  \textcolor{black}{The width of coplanar wave guide resonator is $\SI{10}{\micro m}$, and the gap to ground is $\SI{6}{\micro m}$. Dimensions of airbridges: the width is $\SI{9}{\micro m}$, the length is $\SI{42.6}{\micro m}$, and the height is $\SI{3}{\micro m}$.}
  (c) \textcolor{black}{The data sets of $|S_{21}|$ (shown in left axis for blue colored circle markers) and crosstalk (shown in right axis for red colored cross markers). The center vertical dashed line indicates resonant frequency of resonator 1 $\omega_{r1} = \SI{8.6645}{GHz}$ evaluated at port 3. The vertical dash-dotted line indicate the maximum value of the crosstalk.}
  }
  \label{fig:crosschip}
\end{figure}

\begin{figure}[!b]
  \includegraphics[width=8.4cm]{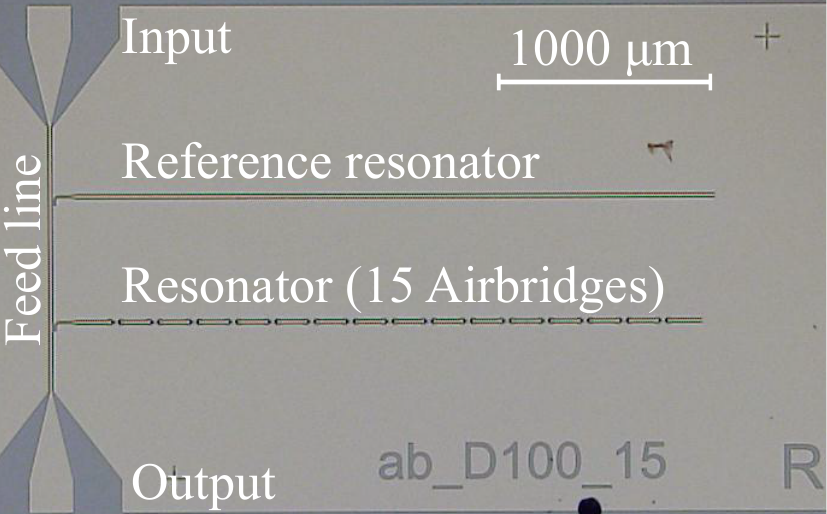}
  \caption{
  \textcolor{black}{
  The measured chip for the resonator with 15 airbridges (top) and the reference resonator (bottom), which are capacitively coupled to feed line.
  They are made out of Nb film on non-doped Si wafer. }
  }
  \label{fig:qfchip}
\end{figure}

\textcolor{black}{The crosstalk between two crossed resonator lines is also evaluated using another chip shown in Fig.~\ref{fig:crosschip}(a). 
A feed line crosses a resonator, vertically, using an airbridge [Fig. 5(b)]. The frequency of resonator $\omega_{r1}$ was measured by port 3. 
We subsequently measured the crosstalk between the feed line and the resonator around the resonant frequency $\omega_{r1}$. The crosstalk is due to the airbridge structure which connect center signal line of the resonator across the feed line. 
A reference continuous microwave signal was applied through the feed line from input port 1 to output port 2 in Fig.~\ref{fig:crosschip}(a). 
Then, the signal was absorbed at resonant frequency $\omega_{r1}$ of the airbridge resonator and it resulted as a small dip. 
In Fig.~\ref{fig:crosschip}(c), the normalized, measured data $|S_{21}|$ with dip is shown (blue colored circle-markers), and the crosstalk defined by $20\log_{10}(1-|S_{21}|) \,\mathrm{dB}$ is also shown (red colored cross-markers). 
The result shows the crosstalk due to the crossing airbridge was $\SI{-49}{dB}$ at most when frequencies are on resonant. }

\textcolor{black}{Therefore, to make the pseudo-2D interconnection network by airbridges, we should assign all the crossed resonators in the network with different frequencies, detuned more than the measured crosstalk bandwidth of $\SI{10}{MHz}$. This would suppress the effective crosstalk to an extremely small value, even much smaller than the characteristic background damping in a typical microwave measurement system.}

To conclude, we proposed a novel scalable architecture of a superconducting quantum circuit for the surface codes, where the standard planar 2D wirings can be adopted for external wirings, with the help of an airbridge-incorporated inter-qubit pseudo-2D resonator network. 
We also carried out the feasibility experimental study of the pseudo-2D resonator network, and showed that there is no fundamental difficulties in realizing it. 
\textcolor{black}{Our result seems to indicate that it may be possible to build a fault-tolerant, large-scale quantum computer by simple monolithic integration technologies. 
We are planning to construct a small scale circuit to further examine and explore the possibility.
}

\begin{figure*}[!t]
  \includegraphics[width=18cm]{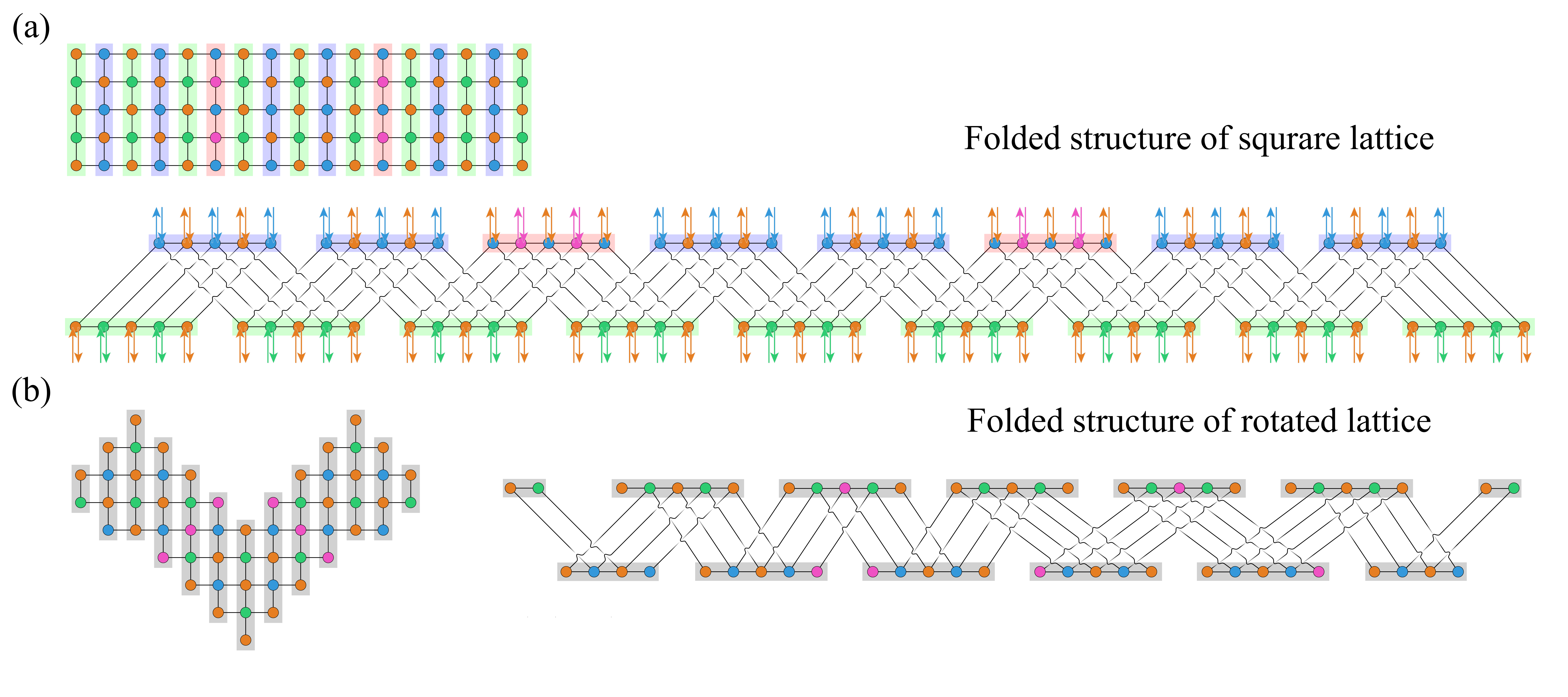}
  \caption{
  \textcolor{black}{
  (a) LNN of the square patch logical qubit and its folded structure for three logical qubits with $d=3$ of the surcface code. 
  (b) LNN of the rotetd patch logical qubit and its folded structure for three logical qubits with $d=3$ of the surcface code. }
  }
  \label{fig:compare}
\end{figure*}

\begin{acknowledgments}
We thank F. Nori, O. V. Astafiev, A. Tomonaga and D. Zhang for discussions.
\textcolor{black}{We also thank M. Hidaka to provide wafers with Nb film.}
This work was supported by CREST, JST. (Grant No. JPMJCR1676), the New Energy and Industrial Technology Development Organization (NEDO), and ImPACT Program of Council for Science, Technology and Innovation (Cabinet Office, Government of Japan).
\end{acknowledgments}

\section*{Author contributions}
H.M., K.S., S.J.D. and J.S.T. designed the architecture.
H.M., R.W. and Y.N. designed the samples, R.W. and Y.N. fabricated the samples.
H.M. and Y.Z. carried out the experiments and analysed the data.
K.S. carried out the numerical calculations. H.M., K.S., S.J.D. and J.S.T. wrote the paper with feedback from all authors.
J.S.T. designed and supervised the project.

\section*{Additional information}
{\bf Competing interests:}
The authors declare that there are no competing interests.

{\bf Corresponding author:}
Correspondence and requests for materials should be addressed to J.S.T.

\section{Methods}
\textcolor{black}{
{\bf Extra information on the experiment:} 
We utilized vector network analyzer (VNA) to measure the internal quality factor and crosstalk. To evaluate the internal quality factor of resonators, we prepare a chip shown in Fig.~\ref{fig:qfchip} with 15 airbridges. The spectrum of resonators are measured, using the input and output ports of the feed line coupled to each resonators.}

{\bf Information on the simulation:}
We modeled a part of our system as two qubits coupled via a damped resonator, so the Hamiltonian is
\begin{equation}\begin{split}
	\mathcal{H}/\hbar =
	& \omega_r a^\dagger a \\ 
	& + \hspace{-0.2cm}\sum_{i=1,2} \biggl[ \textcolor{black}{\omega_i^{01} b_i^\dagger b_i + \frac{\eta_i}{2}} b_i^\dagger b_i (b_i^\dagger b_i - 1) + g_i (a^\dagger b_i + a b_i^\dagger) \biggr],
\end{split}\end{equation}
and this indirect-interaction of qubits (last term) is used for the CZ gate.
The quantum map $\mathcal{E}$ can be derived solving the Lindblad master equation, and then we calculate the average gate (in)fidelity in computational subspace $\ket{\psi_\mathrm{s}}$ between the map $\mathcal{E}$ and an ideal CZ gate map $\mathcal{E}_\mathrm{CZ}$, which is defined as~\cite{wood_quantification_2018},
\begin{equation}
	\overline{F}(\mathcal{E},\mathcal{E}_\mathrm{CZ})
    = \int d\psi_\mathrm{s} \braket{\psi_\mathrm{s} | \mathcal{E}_\mathrm{CZ}^{-1}\circ\mathcal{E}(\psi_\mathrm{s}) | \psi_\mathrm{s}},
\end{equation}
averaged over the Haar measure $d\psi_\mathrm{s}$.
This simulation is performed using Quantum Toolbox in Python (QuTiP)~\cite{johansson_qutip_2013}.

{\bf Using the rotated lattice for logical qubit encoding:}
\textcolor{black}{
In the main text we described the architectural layout using encoded qubits formed from a square lattice of $(2d-1)^2$ physical qubits. 
This can be reduced by utilizing the rotated lattice encoding introduced in Ref.~\cite{Horsman:2012aa}. 
A rotated lattice will reduce the number of physical qubits in a logical block from $(2d-1)^2$ to $2d^2-1$, which for large values of $d$, can result in significant resource savings.  
}

\textcolor{black}{
In terms of the hardware architecture itself, there is no changes that is needed for the underlying hardware. 
In Fig.~\ref{fig:compare} we illustrate how two encoded qubits in the rotated lattice are translated to the bi-linear design. 
Unlike the case when encoded qubits are square patches, the airbridge connections become non uniform. 
However, the maximum number of airbridges within a single resonator does not change between the case of square encoding and rotated encoding. 
Consequently, the design in the main text is completely compatible with using rotated lattice encoding. 
}

{\bf Data availability:}
The data that support the findings of this study are available from the authors on reasonable request, see author contributions for specific data sets.

\textcolor{black}{
{\bf Code availability:}
The numerical data sets presented in Fig.~\ref{fig:qfactor} and Fig.~\ref{fig:crosschip} of this paper are analyzed by MATLAB by H.M., and the numerical data presented in Fig.~\ref{fig:fidelity} of this paper is simulated by H.M. and K.S. The code used to generate this data will be made available to the upon reasonable request.}

\bibliographystyle{apsrev4-1}
\bibliography{bib1,bib2}

\end{document}